\documentclass[preprint,12pt]{elsarticle}
\usepackage{amssymb,amsmath,epsf}
\usepackage{array,threeparttable}

\journal{Nuclear Physics A}

\def\tlab{\mbox{$T_{\rm Lab}$}}

\def\vz{{\mbox{\boldmath$0$}}}

\def\pzp{p_0^{\prime}}

\newcommand{\bp}{{\bf p}}
\newcommand{\bpp}{{\bf p}^{\prime}}
\newcommand{\bk}{{\bf k}}

\def\Spp{^3S_{1}^{+}}
\def\Dpp{^3D_{1}^{+}}

\def\Spp{^3S_1}
\def\Dpp{^3D_1}

\begin{document}

\begin{frontmatter}

%\title{Covariant separable kernel of the deuteron}
\title{Covariant separable interaction for the neutron-proton system in $^3S_1$-$^3D_1$ partial-wave state}

\author[jinr]{S.\,G. Bondarenko}
\ead{bondarenko@jinr.ru}
\author[jinr]{V.\,V. Burov}
\ead{burov@theor.jinr.ru}
\author[hwa]{W-Y. Pauchy Hwang}
\ead{wyhwang@phys.ntu.edu.tw}
\author[jinr]{E.\,P. Rogochaya\corref{cor1}}
\ead{rogoch@theor.jinr.ru}
\address[jinr]{Bogoliubov Laboratory of Theoretical Physics, Joint Institute for Nuclear Research, Dubna, Russia}
\address[hwa]{National Taipei University, Taipei 106, Taiwan}

\cortext[cor1]{Corresponding author. JINR, Joliot-Curie 6, 141980
Dubna, Moscow region, Russia. Tel.: +74962163503; fax: +74962165146}

\begin{abstract}
Within a covariant Bethe-Salpeter approach a rank-six separable
neutron-proton interaction kernel for the triplet coupled $^3S_1$-$^3D_1$ partial-wave state is constructed.
Two different methods of a relativistic generalization of initially nonrelativistic form
factors parametrizing the kernel are considered. The model parameters
are determined by fitting the elastic $^3S_1$ and $^3D_1$ phase shifts and
the triplet scattering length as well as the asymptotic $D/S$ ratio of the deuteron
wave functions and the deuteron binding energy. The $D$-state probability constraints 4-7\% are taken into account. The deuteron
magnetic moment is calculated. The half-off-shell properties are further demonstrated by the Noyes–Kowalski functions. The first test of the
constructed kernel is performed by calculating the deuteron
electrodisintegration at three different kinematic conditions.
\end{abstract}


\end{frontmatter}

\section{Introduction}
The deuteron has been an object of intensive investigations as the
simplest bound neutron-proton system. Throughout more
than 40 years many methods for the description of the deuteron
have been elaborated \cite{Brown:1975us}-\cite{Bondarenko:2002zz}. The
main goal of any approach is the description of the interaction
between two nucleons. It is presented either by potentials
in the Schr$\ddot{{\rm o}}$dinger equation or interaction kernels in the
Bethe-Salpeter (BS) equation. They are either constructed from the
presentation of the interaction as an exchange by various mesons
(realistic potentials \cite{Lacombe:1980dr,Machleidt:1984xu}) or
presented as model functions whose parameters are found from the
description of observables in the elastic neutron-proton ($np$)
scattering (separable
\cite{Crepinsek:1974pu}-\cite{Haidenbauer:1986zza},
phenomenological \cite{Reid:1968sq} models etc). There are also
alternative approaches
\cite{Carbonell:1998rj,krutov:2009,Azhgirey}, relativistic quantum
mechanics, which are based on the Hamiltonian approach.

From the viewpoint of simplicity of performing calculations presentation of
interaction in a separable form is the most convenient instrument \cite{Bondarenko:2002zz}.
That is why there are separable approximations
intended only to reproduce the behavior of the corresponding
realistic potentials and used in calculations instead of more
complicated originals (see, for example, \cite{Lacombe:1980dr} and
\cite{Haidenbauer:1984dz}). However, the construction of presentations of
this type is a complicated problem. The first elaborated models
\cite{Crepinsek:1974pu,Crepinsek:1975vn} were nonrelativistic and,
therefore, they were of little use for the description of reactions
with high-energy particles. In addition, there were problems with
their off-shell behavior, see, for example
\cite{Haftel:1976zz,Giraud:1979gw}. This behavior was adjusted in
subsequent models
\cite{Mathelitsch:1981mr}-\cite{Haidenbauer:1986zza} by fitting to the
corresponding realistic potentials using the Ernst-Shakin-Thaler
method \cite{Ernst:1973zzb}. However, all these models do not
contain the zero component of the momentum of considered particles
which is necessary to construct a covariant model. One of the attempts
of this kind is \cite{Rupp:1989sg} where the relativistically
generalized version of Graz II potential \cite{Mathelitsch:1981mr}
was proposed. It describes the experimental data for the
laboratory energies of the colliding neutron and proton $\tlab$ up
to 0.5\,GeV. However, its application is limited in principle
because of nonintegrability of expressions containing the constructed form
factors \cite{Bondarenko:2002zz} at higher $\tlab$.
The problem can be solved by using the modified form factors
\cite{Schwarz:1980bc}. This idea was developed in
\cite{Bondarenko:2004pn}-\cite{Bondarenko:2008mm} for the
description of uncoupled partial-wave states in the elastic $np$
scattering for $\tlab$ up to 3\,GeV.

In the present paper the separable interaction kernel for the triplet partial-wave
state $^3S_1$-$^3D_1$ is proposed.
%(In what follows, the parity labels would be indicated explicitly.)
The work is a continuation
of the previous one \cite{Bondarenko:2008mm} where the uncoupled
partial-wave states with the total angular
momenta $J=0,1$ were considered. Investigations of various deuteron
characteristics are performed using the elaborated kernel.
Parameters of the model are defined from the calculations of
experimental data for phase shifts taken from the SAID program
(http://gwdac.phys.gwu.edu) and low-energy characteristics. The
off-shell behavior of the kernel is compared with the deuteron
wave function and the Noyes-Kowalski function
\cite{Noyes:1965ib,Kowalski:1965} obtained for some realistic
potentials (here Paris, CD-Bonn) which are very good at low energies.

The paper is organized as follows. In. Sec. 2, the general Bethe-Salpeter
formalism used for the description of the $np$ system is considered.
In Sec. 3, the solution of the BS equation using the separable
presentation of its kernel is discussed. The proposed model is expounded
in Sec. 4. The calculations of introduced parameters and obtained
low-energy deuteron characteristics are presented in Sec. 5.
The review of the results for phase shifts of the elastic $np$
scattering, components of the Noyes-Kowalski function, deuteron
wave function and the conclusions are given in Sec. 6.
\section{Bethe-Salpeter formalism}\label{sect2}
In the relativistic field theory, elastic nucleon-nucleon (NN) scattering
can be described by the scattering matrix $T$ which satisfies the
inhomogeneous Bethe-Salpeter equation \cite{Salpeter:1951sz}.
In momentum space, the BS equation for
the $T$ matrix can be (in terms of the relative four-momenta
$p^\prime$ and $p$ and the total four-momentum $P$) represented as:
\begin{eqnarray}
T(p^{\prime}, p; P) = V(p^{\prime}, p; P) + \frac{i}{4\pi^3}\int
d^4k\, V(p^{\prime}, k; P)\, S_2(k; P)\, T(k, p; P), \label{t00}
\end{eqnarray}
where $V(p^{\prime}, p; P)$ is the interaction kernel and
$S_2(k; P)$ is the free two-particle Green function
$$S_2^{-1}(k; P)=\bigl(\tfrac12\:P\cdot\gamma+{k\cdot\gamma}-m\bigr)^{(1)}
\bigl(\tfrac12\:P\cdot\gamma-{k\cdot\gamma}-m\bigr)^{(2)},$$
$\gamma$ are the Dirac gamma-matrices.
The square of the total momentum $s=(p_1+p_2)^2$ and the
relative momentum $p=(p_1-p_2)/2$ [$p'=(p_1'-p_2')/2$] are
defined via the nucleon momenta
$p_1,~p_2$ [$p_1',~p_2'$] of initial [final] nucleons.

Performing the partial-wave decomposition (see details in
\cite{Bondarenko:2002zz,Bondarenko:2008mm}) of the $T$ matrix and
interaction kernel $V$ we can rewrite the BS equation for the
off-shell partial-wave amplitudes:
\begin{eqnarray}
&&T_{ab}(\pzp, |\bpp|; p_0, |\bp|; s) =
V_{ab}(\pzp, |\bpp|; p_0, |\bp|; s)
\hskip 50mm
\label{t04}\\
&&\hskip 38mm+ \frac{i}{4\pi^3}\sum_{cd}\int\limits_{-\infty}^{+\infty}\!
dk_0\int\limits_0^\infty\! \bk^2 d|\bk|\, V_{ac}(\pzp, |\bpp|; k_0,
|\bk|; s)\nonumber\\
&&\hskip 38mm\times\, S_{cd}(k_0,|\bk|;s)\, T_{db}(k_0,|\bk|;p_0,|\bp|;s). \nonumber
\end{eqnarray}
Here indices $a, b$ etc denote the corresponding partial-wave
state quantum numbers $|aM\rangle\equiv|\pi,\,{}^{2S+1}L_J^{\rho}
M \rangle$ \cite{Kubis:1972zp}, where $S$ is the total spin, $L$
is the orbital angular momentum, and $J$ is the total angular
momentum with the projection $M$; relativistic quantum numbers
$\rho$ and $\pi$ refer to the relative-energy and spatial parity
with respect to the change of sign of the relative energy and
spatial vector, respectively. The two-spinor propagator $S_{ab}$
depends only on $\rho$-spin indices. The quantum number $\rho$
defines the positive- ($\rho=+$) or negative-energy ($\rho=-$)
partial-wave states. In nonrelativistic models of nuclear-nuclear
interactions only positive-energy states are considered; therefore,
$\rho$ is superfluous to be shown as a quantum number. In the
relativistic model the $\rho=-$ states should be described in the
general case. However, since only positive-energy partial states are considered in the paper below $\rho$ is omitted.

Calculating the $T$ matrix we can connect the parameters of the BS kernel $V$
with observables.
For the description of the $T$ matrix we use the following
normalization condition in the
on-mass-shell form for the triplet state:
\begin{eqnarray}
T_{l'l}(s)=\frac{i8\pi}{\sqrt s\sqrt{s-4m^2}}\left(
\begin{array}{cc}
\cos 2\varepsilon_1\,e^{2i\delta_{<}}-1 & i\sin 2\varepsilon_1 \,e^{i(\delta_{<}+\delta_{>})} \\
i\sin 2\varepsilon_1\,e^{i(\delta_{<}+\delta_{>})} & \cos 2\varepsilon_1\,e^{2i\delta_{>}}-1
\end{array}
\right),\label{T_norm_t}
\end{eqnarray}
where $m$ is a nucleon mass and $\varepsilon_1$ is a mixing parameter.
In Eq.(\ref{T_norm_t}) $\delta_<=\delta_{L=J-1}$,
$\delta_{>}=\delta_{L=J+1}$ and $l$ denotes
$^{2S+1}L_J$ states for simplicity.
Expanding the $T$ matrix into a series of $\bar p$-terms,
according to \cite{Bethe:1949yr},
\begin{eqnarray}
\bar p\cot\delta_l(s)=-\frac{1}{a_0^l}+\frac{r_0^l}{2}\bar p^2
+{\cal O}(\bar p^3),\label{low}
\end{eqnarray}
where
\begin{eqnarray}
\bar p\equiv |{\bar \bp|} =
\sqrt{s/4-m^2}=\sqrt{m\tlab/2}\label{pbar}
\end{eqnarray}
is the on-mass-shell momentum, one can derive low-energy parameters, the
scattering length $a_0$ and the effective range $r_0$.

The bound state of the two-particle system appears as a simple pole
in the $T$ matrix at $s=M_d^2$, with $M_d$ being a mass of a bound state,
in our case it is a deuteron.
Thus, the Bethe-Salpeter equation for the BS amplitude
$\Phi$ of the two-nucleon system with the total momentum $J$ and its projection
${M}$ has the following form:
\begin{eqnarray}
\Phi^{J{M}}(p;P)= \frac{i}{(2\pi)^4} S_2(p; P) \int
{d^4k}V(p,k;P)\Phi^{J{M}}(k;P), \label{BS_Phi}
\end{eqnarray}
The partial-wave decomposed amplitude $\Phi$
can be written in the rest frame of the particles
through the generalized spherical harmonic ${\cal Y}$ and the radial $\phi$ part as:
\begin{eqnarray}
\Phi^{J{M}}_{\alpha\beta}(p; P_{(0)}) = \sum_{a}
({\cal Y}_{aM}({\bp})U_C)_{\alpha\beta}\ \phi_{a}(p_0,|\bp|;s), \label{phi00}
\end{eqnarray}
where $P_{(0)}=(M_d,\vz)$ is the total momentum of the NN system in its rest frame.
Here $U_C$ is the charge conjugation matrix.
%$L$ is an orbital momentum and $S$ is a spin of the partial state,
%the quantum number $\rho=+1$ corresponds to the particle,
%$\rho=-1$ - to antiparticle.
In the numerical calculations instead of the amplitude it is more convenient
to use the vertex function $\Gamma$ which is connected with the BS
amplitude:
\begin{eqnarray}
\Phi_{J{M}}(p;P)= S_2(p; P) \Gamma_{J{M}}(p;P),
\label{BS_vf}
\end{eqnarray}
so the radial parts of the BS amplitude $\phi$ and the vertex function $g$
are connected by the relation:
\begin{eqnarray}
\phi_{a}(p_0,|\bp|)=\sum_{b}S_{ab}(p_0,|\bp|;s)g_{b}(p_0,|\bp|). \label{amp_vf}
\end{eqnarray}

To solve the equations for the $T$ matrix and BS amplitude, one should
use some assumption for the kernel $V$. In our case it is
a separable ansatz.

\section{Separable model}\label{sect3}

We assume that the interaction conserves parity,
%total spin $S$,
total angular momentum  $J$ and its projection, and isotopic spin.
Due to the NN tensor force, the orbital angular momentum
$L$ is not conserved. Moreover, the negative-energy two-nucleon
states are omitted, which leads to the total spin $S$
conservation. The partial-wave-decomposed BS equation is therefore
reduced to the following form:
\begin{eqnarray}
T_{l'l}(\pzp, |\bpp|; p_0, |\bp|; s) =
V_{l'l}(\pzp, |\bpp|; p_0, |\bp|; s)
\hskip 50mm
\label{BS}\\
+ \frac{i}{4\pi^3}\sum_{l''}\int\limits_{-\infty}^{+\infty}\!
dk_0\int\limits_0^\infty\! \bk^2 d|\bk|\, \frac{V_{l'l''}(\pzp,
|\bpp|; k_0,|\bk|; s)\, T_{l''l}(k_0,|\bk|;p_0,|\bp|;s)}
{(\sqrt{s}/2-E_{\bk}+i\epsilon)^2-k_0^2}. \nonumber
\end{eqnarray}

Supposing the separable (rank-$N$) ansatz for $V$:
\begin{eqnarray}
V_{l'l}(\pzp, |\bpp|; p_0, |\bp|; s)=\sum_{i,j=1}^N\lambda_{ij}(s)
g_i^{[l']}(\pzp, |\bpp|)g_j^{[l]}(p_0, |\bp|), \label{V_separ}
\end{eqnarray}
where the form factors $g_j^{[l]}$ represent the model functions, we can
obtain the solution of Eq.(\ref{BS}) in a similar
separable form for the $T$ matrix:
\begin{eqnarray}
T_{l'l}(\pzp, |\bpp|; p_0, |\bp|; s)=
\sum_{i,j=1}^N\tau_{ij}(s)g_i^{[l']}(\pzp, |\bpp|)
g_j^{[l]}(p_0, |\bp|),
\end{eqnarray}
where
\begin{eqnarray}
\tau_{ij}(s)=1/(\lambda_{ij}^{-1}(s)+h_{ij}(s)),
\end{eqnarray}
\begin{eqnarray}
h_{ij}(s)=-\frac{i}{4\pi^3}\sum_{l}\int dk_0\int
\bk^2d|\bk| \frac{g_i^{[l]}(k_0,|\bk|)g_j^{[l]}(k_0,|\bk|)}{(\sqrt
s/2-E_{\bk}+i\epsilon)^2-k_0^2}, \label{H_separ}
\end{eqnarray}
$\lambda_{ij}(s)$ is a matrix of model parameters
with the symmetry property:
\begin{eqnarray}
\lambda_{ij}(s)=\lambda_{ji}(s).
\end{eqnarray}

Using the separable ansatz (\ref{V_separ}) and performing the
partial-wave decomposition for the interaction kernel and vertex function
we can represent the radial part of the latter as follows:
\begin{eqnarray}
g_l(p_0,|\bp|)=\sum_{i,j=1}^N{\lambda_{ij}(s)g_i^{[l]}(p_0,|\bp|)c_j(s)},
\label{phi_separ}
\end{eqnarray}
where $l=S(D)$ corresponds to the $^3S_1(^3D_1)$ wave in the deuteron,
and the integral equation (\ref{BS_Phi}) is reduced to a system
of linear homogeneous equations for the coefficients $c_i(s)$:
\begin{eqnarray}
c_i(s)-\sum_{k,j=1}^N h_{ik}(s)\lambda_{kj}(s)c_{j}(s)=0. \label{C_i}
\end{eqnarray}
The radial part of the BS amplitude has the form (see Eq.(\ref{amp_vf})):
\begin{eqnarray}
\phi_{l}(p_0,|\bp|)= \frac{g_{l}(p_0,|\bp|)}{({M_d}/2-E_{\bp}+i\epsilon)^2-p_0^2}.
\label{amp_vf_d}
\end{eqnarray}

The form factors $g_i^{[l]}$ used in the separable representation
(\ref{V_separ}) are obtained by a relativistic
generalization of the initially nonrelativistic Yamaguchi-type
functions depending on the three-dimensional squared momentum $|\bp|$.
Methods of covariant relativistic
generalizations and the constructed
form factors are discussed in the next section.
\section{Construction of the covariant separable kernel}\label{sect6}
We consider two types of a relativistic generalization of
nonrelativistic Yamaguchi-type form factors \cite{Yamaguchi:1954mp} performing
the following changes:
\begin{eqnarray}
&&\bp^2 \to -p^2 = -p_0^2 + \bp^2 \label{p2p}\\
{\rm or}~~&&\bp^2 \to Q^2:~Q = p - \frac{P\cdot p}{s} P \label{p2Q}
\end{eqnarray}
and denote the resulting modified form factors of the constructed
rank-six interaction kernels by MY$6$ (\ref{p2p}), MYQ$6$ (\ref{p2Q}).
So high rank is necessary to describe simultaneously phase shifts for two
waves and bound state characteristics in addition to usual
low-energy parameters. The form factors
have the following form:
\begin{eqnarray}
&&g^{[S]}_1(p)=\frac{(p_{c1}-p_0^2+\bp^2)}
{(p_0^2-\bp^2-\beta_{1}^2)^2+\alpha_{1}^4},
\label{sgfunc}\\
%\end{eqnarray}
%\begin{eqnarray}
&&g^{[S]}_2(p)=\frac{(p_0^2-\bp^2)(p_{c2}-p_0^2+\bp^2)^2}
{((p_0^2-\bp^2-\beta_{2}^2)^2+\alpha_{2}^4)^2},
\nonumber\\
&&g^{[S]}_3(p)=\frac{(p_0^2-\bp^2)^3(p_{c3}-p_0^2+\bp^2)^2}
{((p_0^2-\bp^2-\beta_{3}^2)^2+\alpha_{3}^4)^3},\nonumber
\end{eqnarray}
\begin{eqnarray}
g^{[S]}_4 = g^{[S]}_5= g^{[S]}_6=g^{[D]}_1 = g^{[D]}_2= g^{[D]}_3=0,
\end{eqnarray}
\begin{eqnarray}
&&g^{[D]}_4(p)=\frac{-(p_0^2-\bp^2)(p_{c4}-p_0^2+\bp^2)^2}
{((p_0^2-\bp^2-\beta_{41}^2)^2+\alpha_{41}^4)
((p_0^2-\bp^2-\beta_{42}^2)^2+\alpha_{42}^4)},
\label{dgfunc}\\
%\end{eqnarray}
%\begin{eqnarray}
&&g^{[D]}_5(p)=\frac{-(p_0^2-\bp^2)}{(p_0^2-\bp^2-\beta_{5}^2)^2+\alpha_{5}^4},
\nonumber\\
&&g^{[D]}_6(p)=\frac{(p_0^2-\bp^2)^4(p_{c6}-p_0^2+\bp^2)}
{((p_0^2-\bp^2-\beta_{61}^2)^2+\alpha_{61}^4)^2((p_0^2-\bp^2-\beta_{62}^2)^2+\alpha_{62}^4)}.
\nonumber
\end{eqnarray}
The functions for the model approximation MYQ6 can be obtained from (\ref{sgfunc})-(\ref{dgfunc})
by the change $p^2\rightarrow Q^2$. The detailed discussion of properties
of the presented form factors can be found in
\cite{Bondarenko:2008fp}-\cite{Bondarenko:2008mm}.

Due to the structure of the separable presentation, Eqs.(\ref{sgfunc})-(\ref{dgfunc}), the matrix $H=\{h_{ij}\}$
(\ref{H_separ}) can be written as
\begin{eqnarray}
H(s)=\left(
\begin{array}{cccccc}
h_{11}(s) & h_{12}(s) & h_{13}(s) & 0         & 0         & 0         \\
h_{12}(s) & h_{22}(s) & h_{23}(s) & 0         & 0         & 0         \\
h_{13}(s) & h_{23}(s) & h_{33}(s) & 0         & 0         & 0         \\
0         & 0         & 0         & h_{44}(s) & h_{45}(s) & h_{46}(s) \\
0         & 0         & 0         & h_{45}(s) & h_{55}(s) & h_{56}(s) \\
0         & 0         & 0         & h_{46}(s) & h_{56}(s) & h_{66}(s)
\end{array}
\right).
\end{eqnarray}

The solution of the BS equation for the radial parts of the vertex function
can be expressed through the introduced form factors with the help of
Eqs.(\ref{amp_vf}), (\ref{phi_separ}), (\ref{C_i}).
To fix the coefficients $c_i$ which are solutions of a system
of linear homogeneous equations, we use the normalization condition for
the $\Spp$ and $\Dpp$ states:
\begin{eqnarray}
p_S + p_D = 1,
\end{eqnarray}
where the pseudoprobabilities of these waves are introduced:
\begin{eqnarray}
p_l = \frac{i}{2M_d(2\pi)^4}\int dp_0 \int \bp^2 d|\bp| \,
\frac{(E_{\bp}-M_d/2) [g_{l}(p_0,|\bp|)]^2}
{\left((M_d/2-E_{\bp}+i\epsilon)^2-p_0^2\right)^2}, \label{norm}
\end{eqnarray}
The resulting coefficients $c_i$ are presented in Table \ref{coefs}.

The half-off-shell behavior of the interaction is controlled by the
Noyes-Ko\-wal\-ski function \cite{Noyes:1965ib,Kowalski:1965} in the pair
rest frame (c.m.) where the corresponding partial-wave decomposition
of the $T$ matrix \cite{Bondarenko:2002zz} is performed:
\begin{eqnarray}
f_{l'l}(p;{\bar p})=\frac{T_{l'l}(0,|\bp|;0,|\bar{\bp}|;s)}{T_{l'l}(0,|\bar{\bp}|;0,|\bar{\bp}|;s)}
\label{NKhosf}
\end{eqnarray}
and compared with the Paris \cite{Lacombe:1980dr} potential and
relativistic Graz II model \cite{Rupp:1989sg}. Here $\bar{\bp}$
(\ref{pbar}) is the on-shell and $\bp$ is the off-shell momenta,
respectively. It should be noted that there are three different
variants of the Graz II separable presentation of the BS kernel differing by the $D$
wave probability. We choose the variant with $p_D=5\%$.

We also constrain the parameters of the separable interaction so
that our model reproduces the deuteron asymptotic $D/S$ ratio
\begin{eqnarray}
\rho_{D/S}=\frac{g_{D}(0,p^*)}{g_{S}(0,p^*)},
\end{eqnarray}
where ${p^*}^2=-mE_d$. Finally, the deuteron magnetic moment
$\mu_d$ is not included in the fit and is calculated as it is.
\begin{center}
\begin{table}[ht]
\caption{Coefficients $c_i$.} \centering
\begin{tabular}{lcc}
\hline\hline
           & MY6 & MYQ6 \\
\hline
$c_1$        &  0.0486061704  &  0.0704040072  \\
$c_2$        &  0.00225058402 &  0.00453017913 \\
$c_3$        & -0.00671506834 & -0.171434026   \\
$c_4$        & -0.019764894   &  0.0080399654  \\
$c_5$        &  0.00277839389 &  0.00618413506 \\
$c_6$        & -0.304144129   &  0.0682392337  \\
\hline\hline
\end{tabular}\label{coefs}
\end{table}
\end{center}
\section{Calculations and results}\label{sect7}
Using the $np$ scattering data we analyze the parameters
of the constructed separable models.
A pole in the $T$ matrix at the mass of the bound state $M_d$:
\begin{eqnarray}
\det|\tau^{-1}_{ij}(s=M_d^2)|=0\label{tauz}
\end{eqnarray}
is taken into account by introducing the additional parameter $m_0$:
\begin{eqnarray}
\lambda_{ij}(s)=\frac{\bar\lambda_{ij}}{s-m_0^2}~.
\end{eqnarray}
The parameter $m_0$ is chosen to satisfy the following condition:
\begin{eqnarray}
\bar\lambda^{-1}_{ij}(s-m_0^2)+h_{ij}(M_d^2)=0,\label{bound_couple}
\end{eqnarray}
where $M_d=(2m-E_d)$ and $E_d$ is the energy of the deuteron.

The calculation of the parameters is performed by using
Eqs.(\ref{T_norm_t}),(\ref{low}) and expressions given
to reproduce experimental values for all
available data from the SAID program (http://gwdac.phys.gwu.edu)
for phase shifts. The low-energy scattering parameters are
taken from~\cite{Dumbrajs:1983jd}.

The calculations are performed using
the Wick rotation \cite{Lee:1969fy}. All integrals are calculated numerically with
the technique elaborated in \cite{Fleischer:1975}.

The introduced free parameters are found from the
minimization of the $\chi^2$ function.
%The probability of the $D$ wave $p_D$ and
The asymptotic ratio
$\rho_{D/S}$
% and the mixing parameter $\varepsilon_1$ are
is also taken into account in the minimization procedure. The phase
shifts for $S$ and $D$ waves are both included in one
minimization function:
\begin{eqnarray}
&&\chi^2=\nonumber\\
&& \sum\limits_{i=1}^{n}(\delta^{\rm
exp}_<(s_i)-\delta_<(s_i))^2/(\Delta\delta^{\rm exp}_<(s_i))^2
+(\delta^{\rm
exp}_>(s_i)-\delta_>(s_i))^2/(\Delta\delta^{\rm exp}_>(s_i))^2
\nonumber\\
&&
%+(\varepsilon_1^{\rm exp}(s_i)-\varepsilon_1(s_i))^2/(\Delta\varepsilon_1^{\rm exp}(s_i))^2
+(a^{\rm exp}-a)^2/(\Delta a^{\rm exp})^2
%+(p_D^{\rm exp}-p_D)^2/(\Delta p_D^{\rm exp})^2\nonumber\\
+(\rho_{D/S}^{\rm exp}-\rho_{D/S})^2/(\Delta \rho_{D/S}^{\rm exp})^2.
\label{mini_sd}
\end{eqnarray}
Here $n$ is a number of experimental points.

The effective range $r_0$
is calculated via the obtained
parameters and compared with the experimental value $r_0^{\rm exp}$.

The calculated parameters of the separable presentations MY6 and MYQ6 are listed in
Tables \ref{my6} and \ref{myq6}, respectively. In Table \ref{lep}, the
calculated low-energy scattering parameters and deuteron
characteristics are compared with the corresponding experimental
values and other models (Graz II \cite{Rupp:1989sg}, Paris
\cite{Lacombe:1980dr}, CD-Bonn \cite{Machleidt:2000ge}).

In Figs.\ref{3s1} and \ref{3d1}, the results of the phase shift
calculations are compared with experimental data and, in addition
to the afore-said theoretical models, with the empirical
SP07 SAID solution \cite{Arndt:2007qn}. The mixing parameter is pictured in
Fig.\ref{eps}. As an example of comparison with a separable model we take here
the nonrelativistic Graz II
\cite{Mathelitsch:1981mr} potential model (denoted by Graz II (NR)).
The comparison with the
relativistic interaction kernel Graz II \cite{Rupp:1989sg} is not presented because using it the phase
shifts and the mixing parameter cannot be calculated in the whole energy
range where these observables are known. As it was discussed in
\cite{{Bondarenko:2002zz},Bondarenko:2008mm}, in this case when $\tlab$ exceeds
some limit value depending on the parameters in separable form factors it is
impossible to perform numerical calculations in principle, whereas our aim is
to compare our MY6 and MYQ6 with results of other models in a wide energy range.
Wherever it is possible to use the relativistic Graz II at high energies we make a comparison with it.

The obtained components of the Noyes-Kowalski function (\ref{NKhosf}) are
presented in Figs.\ref{t00_}-\ref{t22_}. The $S$- and $D$-state
wave functions $\phi({\bar p_0},\bp)$ (\ref{amp_vf_d}), where ${\bar
p_0}=M_d/2-E_\bp$ \cite{Bondarenko:1998fh}, are given in
Figs.\ref{swave} and \ref{dwave}, respectively, and compared with
the corresponding relativistic and nonrelativistic Graz II models
and the Paris potential \cite{Lacombe:1980dr}.

As an illustration of the behavior of the elaborated separable
models when reactions with the deuteron are considered, in
Figs.\ref{cs1}-\ref{cs3}, the results of calculations of cross
sections for the deuteron electrodisintegration within the
relativistic plane-wave impulse approximation are presented. We
compare the obtained results for the differential cross section
with those obtained earlier \cite{Bondarenko:2006zq} using the
relativistic Graz II model \cite{Rupp:1989sg} for three different
kinematic cases \cite{Bussiere:1981mv,TurckChieze:1984fy}.
\begin{center}
\begin{threeparttable}[ht]
\caption{Parameters of the rank-six separable model with modified (\ref{p2p})
Yamaguchi functions MY6.} \centering
\begin{tabular}{lcclc}
\hline\hline
\multicolumn{5}{c}{MY6} \\
%                                &           & MY6  &                         &             \\
\hline $\bar\lambda_{11}^{{\phantom{1}}^{\phantom{1}}}$    (GeV$^2$) & -126.823 &      & $\beta_{1}$(GeV)      & 0.1000189   \\
$\bar\lambda_{12}$    (GeV$^2$) & -1627.106 &      &  $\beta_{2}$(GeV)       &  1.2089089   \\
$\bar\lambda_{13}$    (GeV$^2$) & -78.78723 &      &  $\beta_{3}$(GeV)       &  0.3884728   \\
$\bar\lambda_{14}$    (GeV$^2$) &  1255.789 &      &  $\beta_{41}$(GeV)      &  0.1617235   \\
$\bar\lambda_{15}$    (GeV$^2$) &  1920.741 &      &  $\beta_{42}$(GeV)      &  1.0569099   \\
$\bar\lambda_{16}$    (GeV$^2$) &  23.25852 &      &  $\beta_{5}$(GeV)       &  0.5975024   \\
$\bar\lambda_{22}$    (GeV$^2$) & -1507.037 &      &  $\beta_{61}$(GeV)      &  0.1000189   \\
$\bar\lambda_{23}$    (GeV$^2$) & -202.4665 &      &  $\beta_{62}$(GeV)      &  0.2562457   \\
$\bar\lambda_{24}$    (GeV$^2$) & -1211.809 &      &  $\alpha_{1}$(GeV)      &  1.7190386   \\
$\bar\lambda_{25}$    (GeV$^2$) &  19296.73 &      &  $\alpha_{2}$(GeV)      &  1.1342682   \\
$\bar\lambda_{26}$    (GeV$^2$) & -19.71478 &      &  $\alpha_{3}$(GeV)      &  0.7779747   \\
$\bar\lambda_{33}$    (GeV$^2$) & -4.911057 &      &  $\alpha_{41}$(GeV)     &  0.1143112   \\
$\bar\lambda_{34}$    (GeV$^2$) &  52.90785 &      &  $\alpha_{42}$(GeV)     &  1.9584773   \\
$\bar\lambda_{35}$    (GeV$^2$) & -557.975  &      &  $\alpha_{5}$(GeV)      &  10.71719    \\
$\bar\lambda_{36}$    (GeV$^2$) & -5.781583 &      &  $\alpha_{61}$(GeV)     &  0.1743705   \\
$\bar\lambda_{44}$    (GeV$^2$) &  2388.451 &      &  $\alpha_{62}$(GeV)     &  0.3825411   \\
$\bar\lambda_{45}$    (GeV$^2$) &  1481.914 &      &  $p_{c1}$(GeV$^2$)      & -3.947706    \\
$\bar\lambda_{46}$    (GeV$^2$) &  23.63605 &      &  $p_{c2}$(GeV$^2$)      & -29.997902   \\
$\bar\lambda_{55}$    (GeV$^2$) & -47615.28 &      &  $p_{c3}$(GeV$^2$)      &  3.9076391   \\
$\bar\lambda_{56}$    (GeV$^2$) &  314.5085 &      &  $p_{c4}$(GeV$^2$)      &  0.5632583   \\
$\bar\lambda_{66}$    (GeV$^2$) &  1.135512 &      &  $p_{c6}$(GeV$^2$)      &  0.278038    \\
$m_0$                 (GeV)     &  1.350753 &      &                         &              \\
\hline\hline
\end{tabular}\label{my6}
\end{threeparttable}
\end{center}
\begin{center}
\begin{threeparttable}[ht]
\caption{Parameters of the rank-six separable model with modified (\ref{p2Q})
Yamaguchi functions MYQ6.} \centering
\begin{tabular}{lcclc}
\hline\hline
\multicolumn{5}{c}{MYQ6} \\
%                                &           & MYQ6  &                         &             \\
\hline $\bar\lambda_{11}^{{\phantom{1}}^{\phantom{1}}}$    (GeV$^2$) & -10.574 &      & $\beta_{1}$(GeV)      & 0.100019    \\
$\bar\lambda_{12}$    (GeV$^2$) & -3498.482 &      &  $\beta_{2}$(GeV)       &  0.5978559    \\
$\bar\lambda_{13}$    (GeV$^2$) &  2.031008 &      &  $\beta_{3}$(GeV)       &  0.431291     \\
$\bar\lambda_{14}$    (GeV$^2$) &  98.12    &      &  $\beta_{41}$(GeV)      &  0.210503     \\
$\bar\lambda_{15}$    (GeV$^2$) &  31.318   &      &  $\beta_{42}$(GeV)      &  0.1001783    \\
$\bar\lambda_{16}$    (GeV$^2$) & -30.813   &      &  $\beta_{5}$(GeV)       &  5.5725       \\
$\bar\lambda_{22}$    (GeV$^2$) &  10548.08 &      &  $\beta_{61}$(GeV)      &  0.1120488    \\
$\bar\lambda_{23}$    (GeV$^2$) & -76.928   &      &  $\beta_{62}$(GeV)      &  0.4489458    \\
$\bar\lambda_{24}$    (GeV$^2$) & -3720.413 &      &  $\alpha_{1}$(GeV)      &  1.468872     \\
$\bar\lambda_{25}$    (GeV$^2$) &  37427.57 &      &  $\alpha_{2}$(GeV)      &  1.09121      \\
$\bar\lambda_{26}$    (GeV$^2$) & -208.697  &      &  $\alpha_{3}$(GeV)      &  0.7449097    \\
$\bar\lambda_{33}$    (GeV$^2$) &  0.77     &      &  $\alpha_{41}$(GeV)     &  0.1          \\
$\bar\lambda_{34}$    (GeV$^2$) & -22.51814 &      &  $\alpha_{42}$(GeV)     &  1.945662     \\
$\bar\lambda_{35}$    (GeV$^2$) & -84.478   &      &  $\alpha_{5}$(GeV)      &  4.08         \\
$\bar\lambda_{36}$    (GeV$^2$) &  2.1435   &      &  $\alpha_{61}$(GeV)     &  0.2058123    \\
$\bar\lambda_{44}$    (GeV$^2$) &  895.2475 &      &  $\alpha_{62}$(GeV)     &  0.7052189    \\
$\bar\lambda_{45}$    (GeV$^2$) &  286.5565 &      &  $p_{c1}$(GeV$^2$)      & -4.2035       \\
$\bar\lambda_{46}$    (GeV$^2$) &  19.419   &      &  $p_{c2}$(GeV$^2$)      & -13.82        \\
$\bar\lambda_{55}$    (GeV$^2$) & -32435.14 &      &  $p_{c3}$(GeV$^2$)      &  9.574        \\
$\bar\lambda_{56}$    (GeV$^2$) & -705.156  &      &  $p_{c4}$(GeV$^2$)      &  1.152        \\
$\bar\lambda_{66}$    (GeV$^2$) & -21.792   &      &  $p_{c6}$(GeV$^2$)      & -0.4527       \\
$m_0$                 (GeV)     & 1.351914  &      &                         &               \\
\hline\hline
\end{tabular}\label{myq6}
\end{threeparttable}
\end{center}
\begin{center}
%\begin{table}
\begin{threeparttable}[ht]
\caption{The low-energy scattering parameters for the triplet
$^3S_1$-$^3D_1$ state and
the deuteron characteristics.} \centering
\begin{tabular}{lcccccc}
\hline\hline
           &$a_{0t}$    & $r_{0t}$ & $p_D$ & $E_d$        & $\rho_{D/S}$       & $\mu_d$         \\
           &(fm)        & (fm)     & (\%)  & (MeV)        &                    & $(e/2m)$        \\
\hline
MY6     & 5.42       & 1.800    & 4.92  & 2.2246       & 0.0255             & 0.8500\tnote{a} \\
MYQ6    & 5.42       & 1.768    & 4.65  & 2.2246       & 0.0259             & 0.8535\tnote{a} \\
Graz II & 5.42       & 1.779    & 5     & 2.2254       & 0.0269             & 0.8512          \\
Paris   & 5.43       & 1.770    & 5.77  & 2.2249       & 0.0261             & 0.8469          \\
CD-Bonn & 5.4196     & 1.751    & 4.85  & 2.2246       & 0.0256             & 0.8522          \\
Exp.    & 5.424(4)   & 1.759(5) & 4-7   & 2.224644(46) & 0.0256(4)\tnote{b} & 0.8574\tnote{c} \\
\hline\hline
\end{tabular}\label{lep}
\begin{tablenotes}
\item[a] Calculated within the relativistic impulse approximation with the positive-energy partial-wave states only.
\item[b] Ref. \cite{Rodning:1990zz}.
\item[c] Ref.
\cite{Honzawa:1991qn}.
\end{tablenotes}
\end{threeparttable}
%\end{table}
\end{center}
\section{Discussion and conclusions}\label{sect8}
\begin{figure}
\begin{center}
\includegraphics[width=95mm]{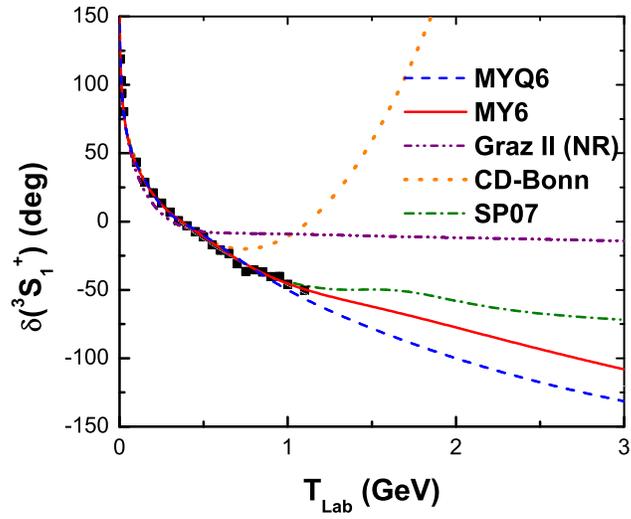}
\caption{{The model phase shifts for the $^3S_1$ and for two relativistic separable kernel cases
MY6 and MYQ6 are compared to those of Graz II (NR) \cite{Mathelitsch:1981mr},
CD-Bonn \cite{Machleidt:2000ge} and of the empirical SP07 SAID solution \cite{Arndt:2007qn}.}}
\label{3s1}
\end{center}
\end{figure}

\begin{figure}
\begin{center}
\includegraphics[width=95mm]{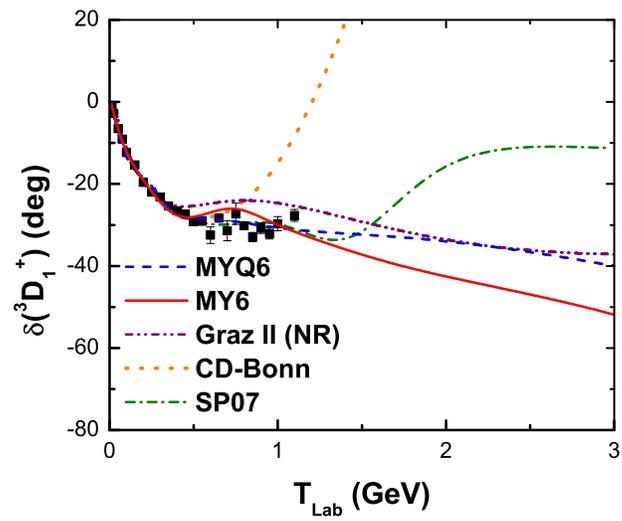}
\caption{{As in Fig.\ref{3s1}, but for the $^3D_1$ wave.}}
\label{3d1}
\end{center}
\end{figure}

The first step in construction of the interaction kernel in a separable form is the description of
on-mass-shell characteristics, like phase shifts, low-energy
characteristics, and the mixing parameter. In
Figs.\ref{3s1} and \ref{3d1}, the results of our calculations of phase
shifts for the $^3S_1$ and $^3D_1$ waves are presented. Figure
\ref{3s1} demonstrates that the $^3S_1$ partial-wave state is well
described by both MY6 and MYQ6 parametrizations for all
experimental data. The Graz II (NR) works for $\tlab\leqslant$0.4\,GeV.
For the $^3D_1$ state MY6 and MYQ6 provide a good description of existing data. Graz II (NR) shows only some correspondence
with the data for $\tlab\leqslant$0.4\,GeV. SP07 is good for all
experimental data. CD-Bonn was constructed for $\tlab\leqslant
350$\,MeV and is perfect in this region. Its behavior means that
other models should be used for higher energies, whereas the interaction kernels
MY6 and MYQ6 allow to perform calculations in this case.

\begin{figure}
\begin{center}
\includegraphics[width=95mm]{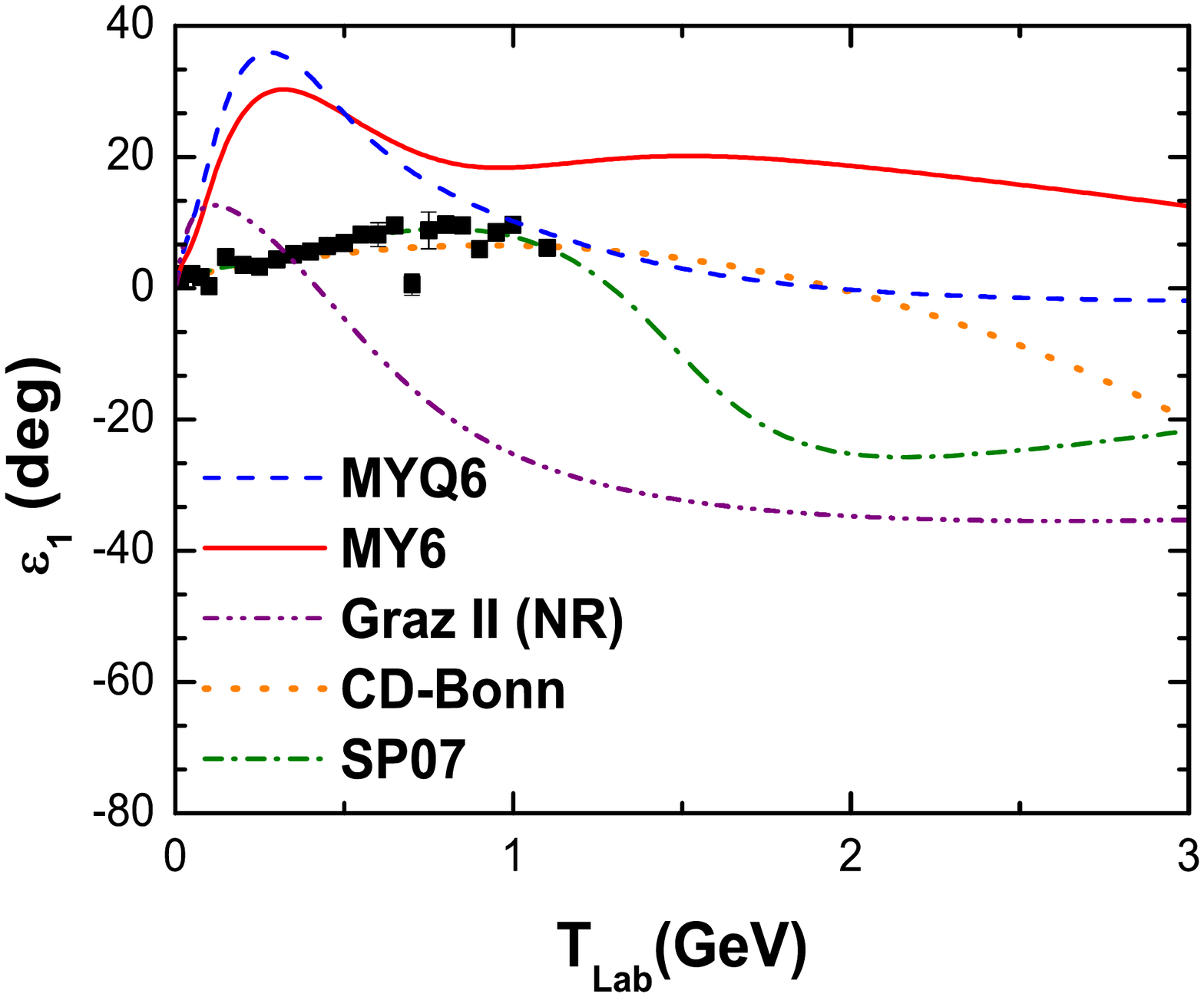}
\caption{{As in Fig.\ref{3s1}, but for the mixing parameter.}}
\label{eps}
\end{center}
\end{figure}

The problem of simultaneous description of phase shifts for the
coupled partial-wave states, the mixing parameter and the
low-energy and deuteron characteristics is worthy of a special
discussion. This subject was considered in detail in
\cite{Mathelitsch:1981mr}. The authors found out that an attempt
to describe well the mixing parameter together with other
observables leads to very bad results for the low-energy and
deuteron parameters. They become too small in comparison with
permissible values, in particular $p_D\lesssim 1\%$. It should be
noted that our investigation of interaction kernels of a rank
$\leqslant 6$ demonstrated that increasing in a rank does not
result in a better description of $\varepsilon_1$. Performing
numerical investigations we discovered the same properties. In
most cases there is no possibility to reproduce phase shifts,
low-energy parameters, and mixing parameter simultaneously. A
certain advance was achieved in the CD-Bonn potential model where
the mixing parameter behavior looks very good for kinetic energies
till about 350\,MeV for Nijmegen group analysis
\cite{Stoks:1993tb}. Although there are SAID group data
\cite{Arndt:2007qn} which differ from Nijmegen group ones this is
the best result for the moment. However, taking into account this
discrepancy in the analysis of the experimental data for the
mixing parameter we restrict ourselves to the description of all
observables but $\varepsilon_1$. At the same time, for
completeness we present the obtained results for $\varepsilon_1$
along with the results of all other discussed models
(Fig.\ref{eps}). It can be seen that both elaborated models do not
work at all like Graz II (NR) potential which also does not give
any agreement with the data. SP07 agrees with the experiment in
the whole energy range.

\begin{figure}
\begin{center}
\includegraphics[width=95mm]{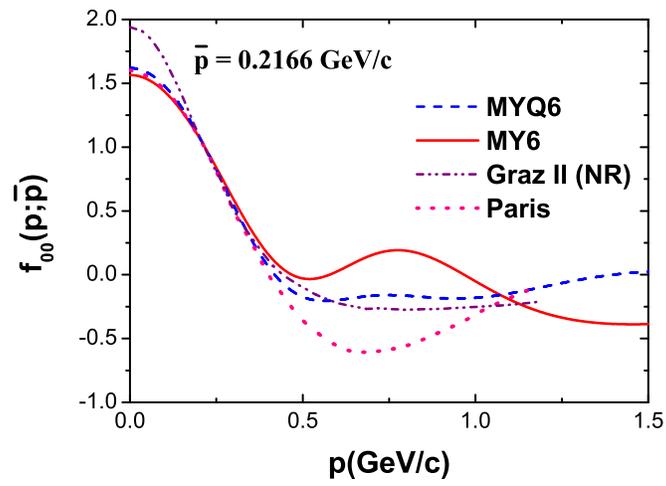}
\caption{{The Noyes-Kowalski function $f_{00}$ in c.m. for two relativistic separable kernel cases
MY6 and MYQ6 is compared to those of the separable Graz II (NR) \cite{Mathelitsch:1981mr} and
realistic Paris \cite{Lacombe:1980dr} potentials.}}
\label{t00_}
\end{center}
\end{figure}

\begin{figure}
\begin{center}
\includegraphics[width=95mm]{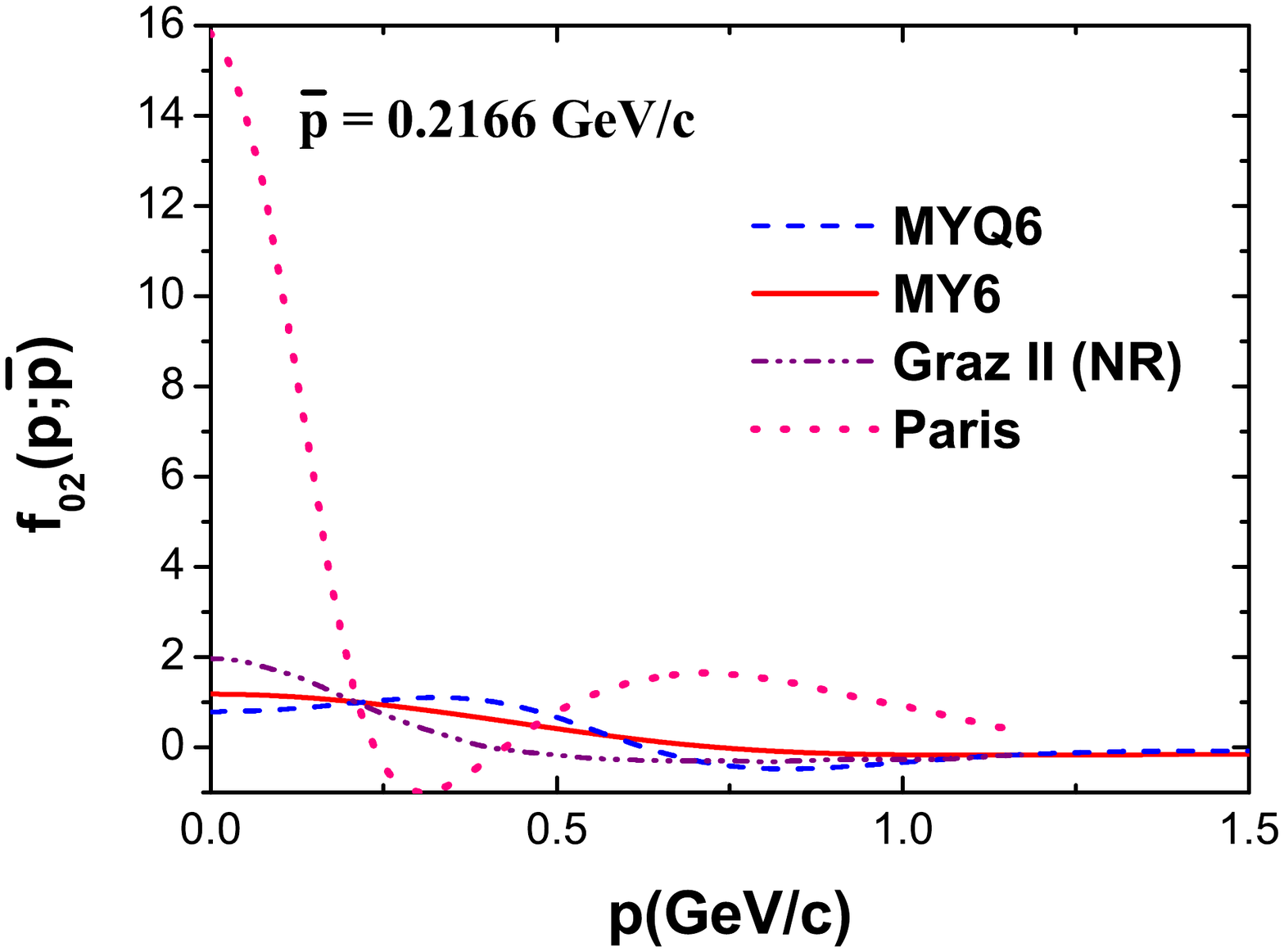}
\caption{{As in Fig.\ref{t00_}, but for the component $f_{02}$ of the Noyes-Kowalski function.}}
\label{t02_}
\end{center}
\end{figure}

\begin{figure}
\begin{center}
\includegraphics[width=95mm]{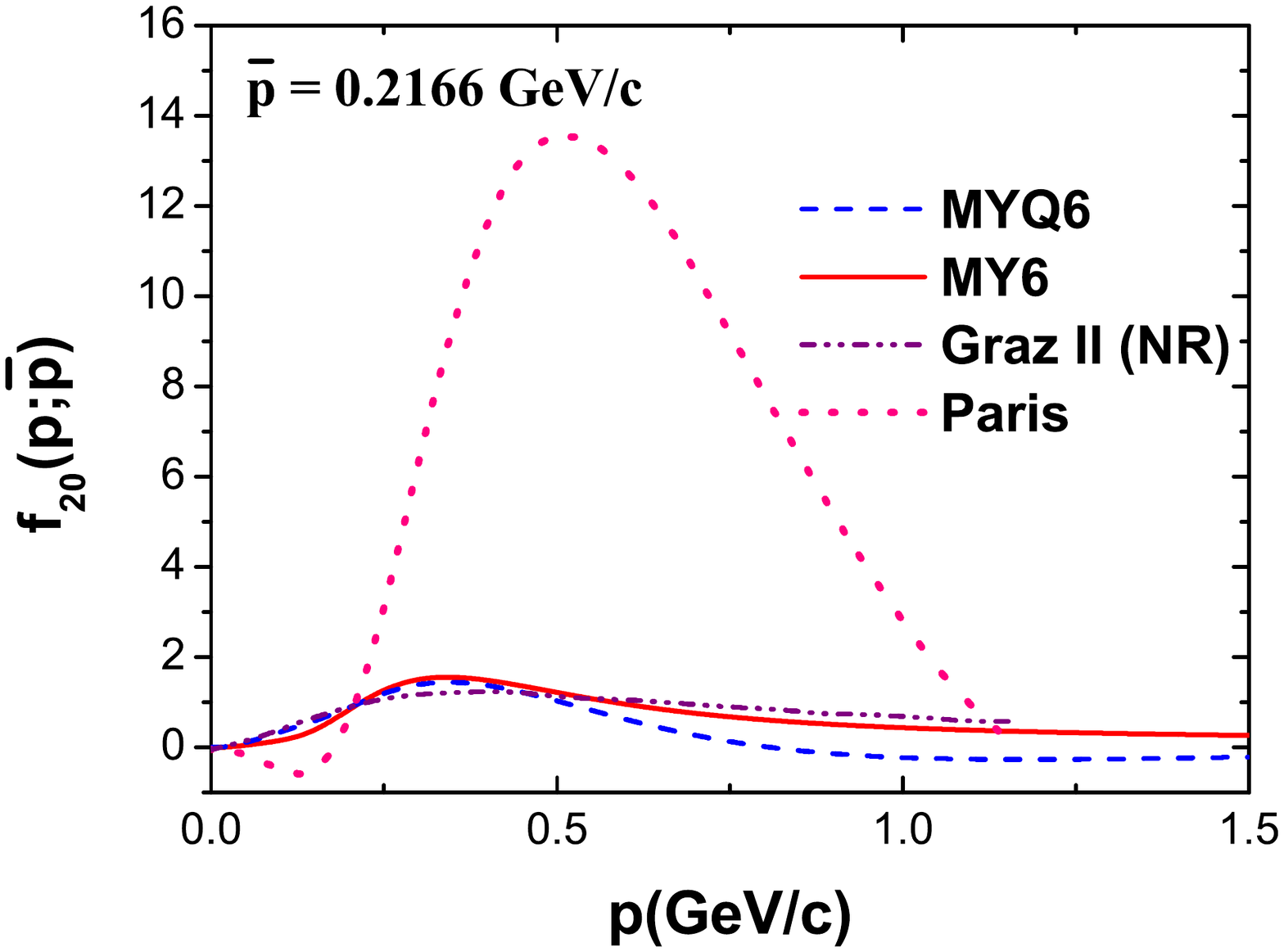}
\caption{{As in Fig.\ref{t00_}, but for the component $f_{20}$ of the Noyes-Kowalski function.}}
\label{t20_}
\end{center}
\end{figure}

\begin{figure}
\begin{center}
\includegraphics[width=95mm]{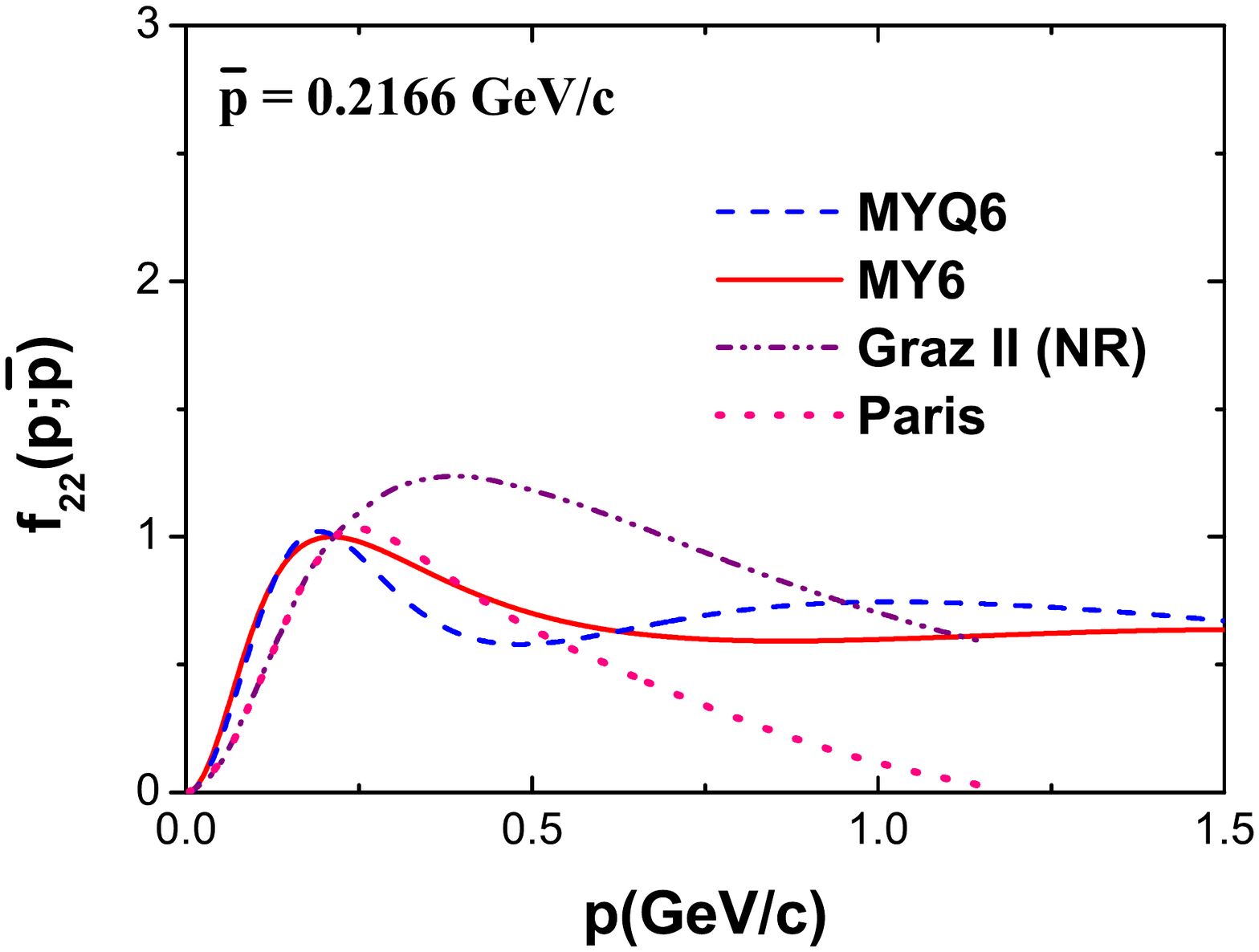}
\caption{{As in Fig.\ref{t00_}, but for the component $f_{22}$ of the Noyes-Kowalski function.}}
\label{t22_}
\end{center}
\end{figure}

The half-off-shell
behavior characterized by the Noyes-Kowalski function (\ref{NKhosf})
was not fitted in any special way, it was simply calculated as it is.
In Figs.\ref{t00_}-\ref{t22_}, all components of this function are presented.
% and compared with Graz II model calculations.
The obtained MY6 and MYQ6 functions are similar but not identical
to the Graz II potential functions. The same can be said about
diagonal components of the function for the realistic Paris
potential, whereas the difference for non-diagonal components is
significant. In addition, MY6 modification is not identical to
MYQ6 one. This difference should affect observables defined by
the half-off-shell behavior of the described interaction,
especially various polarizations \cite{Loiseau:1984eg}.
%However, as it is demonstrated below some results are practically
%identical for observables which depend mostly on the on-shell
%behavior of the interaction.

\begin{figure}
\begin{center}
\includegraphics[width=95mm]{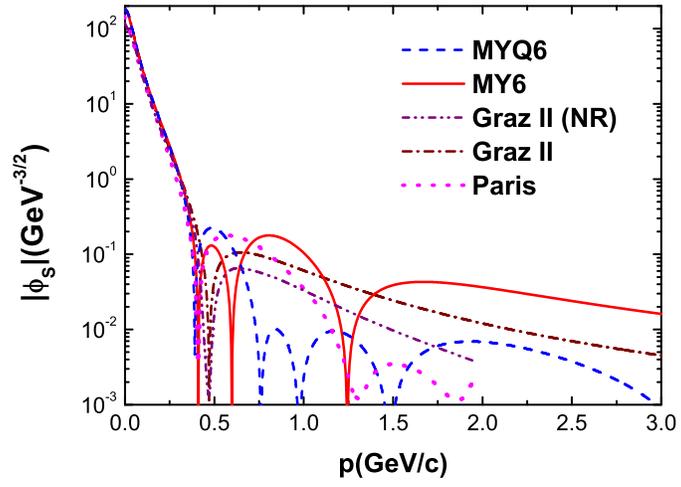}
\caption{{The wave function for the $^3S_1$ partial-wave state in the deuteron
rest frame for the MY6
and MYQ6 models in comparison with those of Graz II (NR) \cite{Mathelitsch:1981mr},
Graz II \cite{Rupp:1989sg} and Paris \cite{Lacombe:1980dr}.}}
\label{swave}
\end{center}
\end{figure}

\begin{figure}
\begin{center}
\includegraphics[width=95mm]{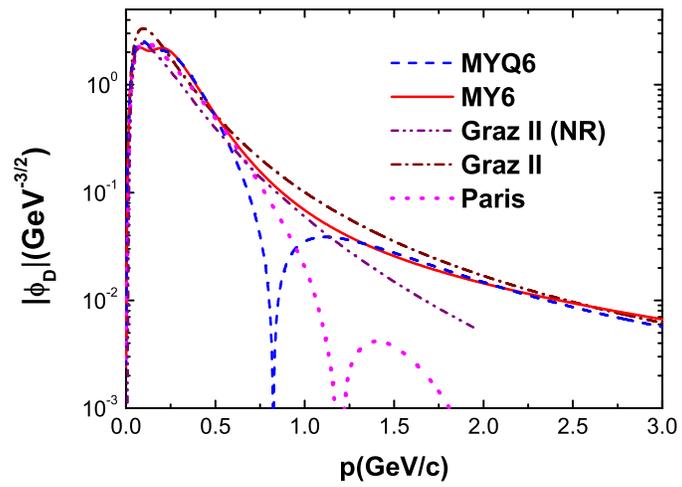}
\caption{{As in Fig.\ref{swave}, but for the $^3D_1$ partial-wave state.}}
\label{dwave}
\end{center}
\end{figure}

As any other phenomenological model ours can describe on-shell
characteristics quite easily. However, the description of the
coupled $^3S_1$-$^3D_1$ channel is not limited only by phase
shifts and low-energy observables. It is also important to look at
properties of the deuteron BS amplitude (wave function). Therefore,
in calculations we take into account features of the wave functions
corresponding to $^3S_1$ and $^3D_1$ parts. In
Figs.\ref{swave} and \ref{dwave}, it is shown that the obtained
functions are similar to other discussed models in the energy
region where their properties play a key role.

\begin{figure}
\begin{center}
\includegraphics[width=95mm]{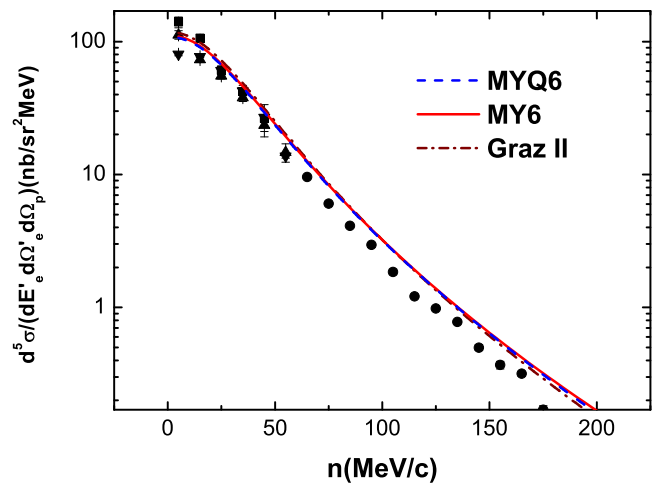}
\caption{{Cross section for the d(e,e$^\prime$p)n reaction, kinematical conditions
set I \cite{Bussiere:1981mv}. $n$ is the recoil momentum of the final neutron. The detailed
discussion of the considered observable can be found in \cite{Bondarenko:2006zq}.}}
\label{cs1}
\end{center}
\end{figure}

\begin{figure}
\begin{center}
\includegraphics[width=95mm]{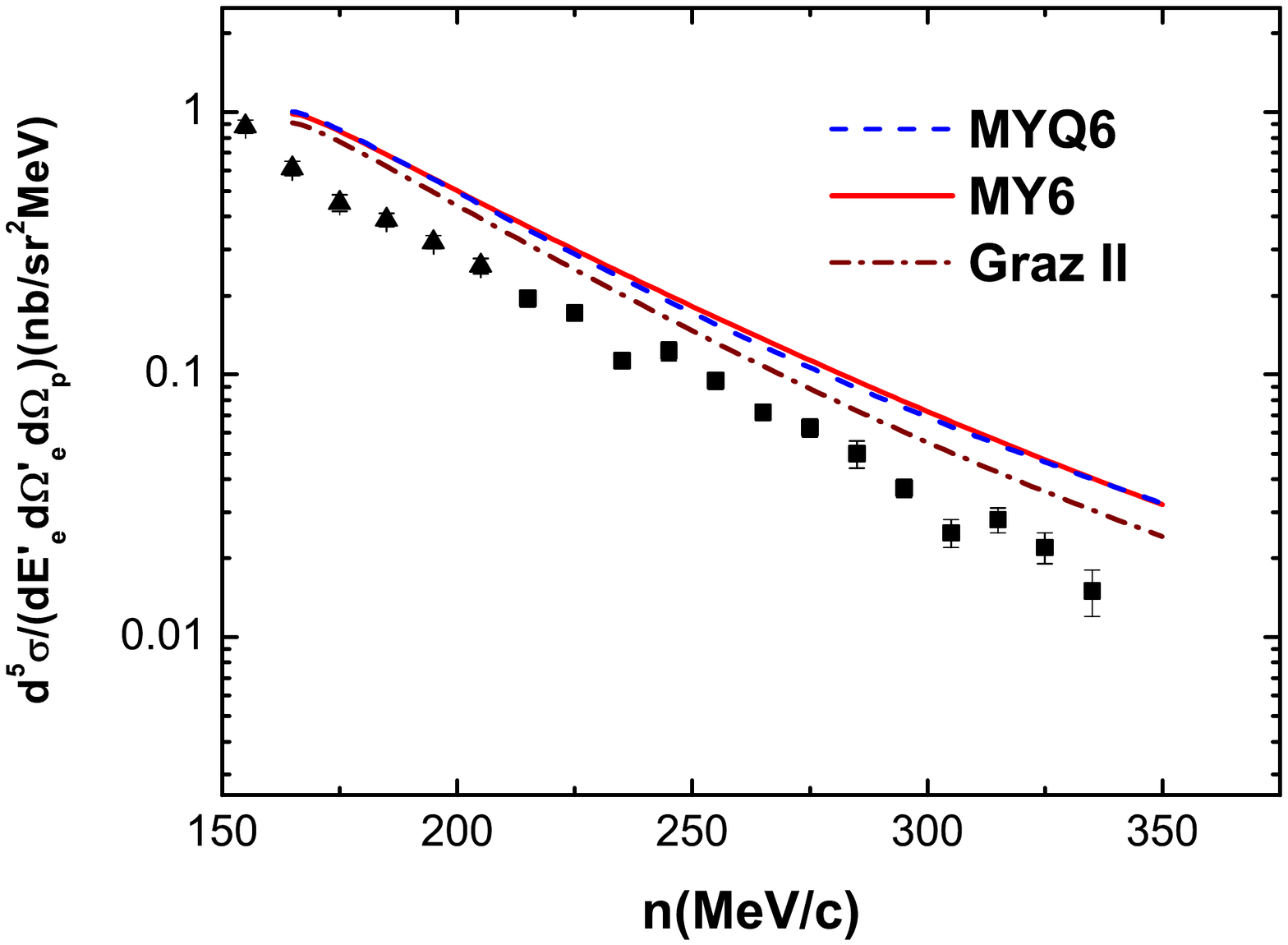}
\caption{{Cross section for the kinematical conditions
set II \cite{Bussiere:1981mv}. See also the caption of Fig.\ref{cs1}.}}
\label{cs2}
\end{center}
\end{figure}

\begin{figure}
\begin{center}
\includegraphics[width=95mm]{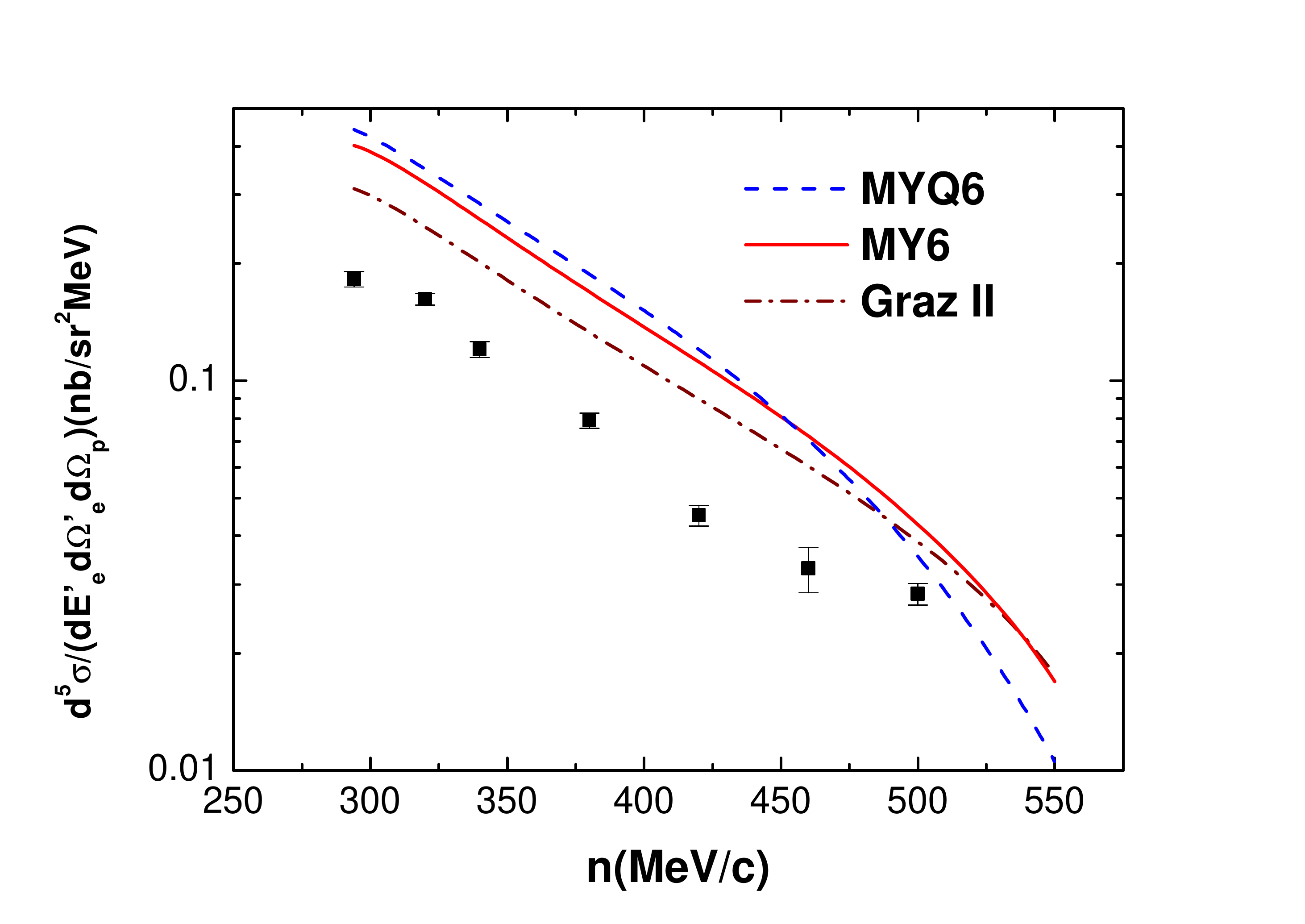}
\caption{{Cross section for the kinematical conditions
\cite{TurckChieze:1984fy}. See also the caption of Fig.\ref{cs1}.}}
\label{cs3}
\end{center}
\end{figure}

As an example of using the interaction kernels constructed in this paper we calculate the differential
cross sections \cite{Bondarenko:2006zq} within the simplest
relativistic plane-wave impulse approximation for various
kinematic conditions \cite{Bussiere:1981mv,TurckChieze:1984fy}
and compare them with the corresponding relativistic Graz II model
\cite{Rupp:1989sg} calculations. From Figs.\ref{cs1}-\ref{cs3},
it can be seen that the presented results begin to differ only
when the influence of the $D$ wave increases. However, to make a
conclusion it is necessary to take into account other effects, like the final state
interaction (FSI) and two-body currents. This was impossible before because of problems with
calculations, as it was discussed in \cite{Bondarenko:2008mm}. Now these
difficulties are obviated and it is planned to perform
calculations of observables with FSI using the elaborated model
and the results of our previous work \cite{Bondarenko:2008mm} in near future.

In conclusion, it can be said that the constructed rank-six interaction kernels
were successfully used for the description of the on-shell and
off-shell characteristics of the triplet $^3S_1$-$^3D_1$ partial-wave state of
the $np$ system and the deuteron. Good agreement with
results of other models which work at low energies for phase
shifts, low-energy parameters, deuteron wave function was
achieved. The demonstrated half-off-shell behavior is similar to the
corresponding Graz II model. The constructed separable model of NN interaction
can be used in
calculations of various reactions with the deuteron, e.g., the
deuteron photo- and electrodisintegration etc. It is also
interesting to investigate the elastic electron-deuteron scattering process using
this new model. However, it is a subject of a separate work.

Additional parameters $\alpha$ provide integrands containing
form factors of the separable presentation to have no poles for $\bp$ component. Therefore, in
particular, using this type of functions for form factors will make numerical
calculations of the electrodisintegration far from the threshold
possible without resorting quasipotential or nonrelativistic
approximations. A comparison with other separable and realistic potential
models allows us to demonstrate merits of separable kernels with $\alpha$-modified
form factors. For example,
the CD-Bonn potential, which was constructed for
$\tlab\leqslant350$\,MeV and works in this energy interval very
well, cannot just be simply extrapolated at higher energies. As for the Graz
II interaction kernel, the increase in energies of the considered particles is
impossible in principle. On the contrary, our model has no these
limitations. We do not describe the mixing parameter; nevertheless,
all low-energy and deuteron characteristics are reproduced well
and the results for phase shifts cover the whole energy region and
are of high quality. Finally, we have no insuperable obstacles in
calculations and there is an opportunity to improve the model in
future.
\section{Acknowledgements}\label{sect10}
We wish to thank Drs. A.A.~Goy, D.V. Shulga, K.Yu.~Kazakov and
Prof. G. Rupp for their interest in the present work and fruitful
discussions. S.G.B. and V.V.B. thank the National Taiwan
University for their warm hospitality.


\begin{thebibliography}{100}

\bibitem{Brown:1975us}
G.E. Brown and A.D. Jackson, The Nucleon-Nucleon Interaction,
\newblock RX-707 (NORDITA), 1974, 151pp; 1975, 387pp.

\bibitem{Gross:1991}
F.~Gross, Relativistic Effects and Relativistic Methods,
in Modern Topics in Electron Scattering, World Scientific,
\newblock 1991, 219pp.

\bibitem{Pascalutsa:2000bs}
V.~Pascalutsa, J.A. Tjon,
\newblock Phys. Rev. C 61 (2000) 054003, nucl-th/0003050.

\bibitem{Mathiot:1984bs}
J.F. Mathiot,
\newblock Nucl. Phys. A 412 (1984) 201.

\bibitem{Wagenbrunn:1995ta}
R.F. Wagenbrunn, W. Plessas,
\newblock Few Body Syst. Suppl. 8 (1995) 181.

\bibitem{Wilbois:1993gn}
T. Wilbois, G. Beck, H. Arenhovel,
\newblock Few Body Syst. 15 (1993) 39.

\bibitem{Carbonell:1998rj}
J. Carbonell, B.~Desplanques, V.A. Karmanov, J.F. Mathiot,
\newblock Phys. Rept. 300 (1998) 215, nucl-th/9804029.

\bibitem{Gari:1976kj} M. Gari, H. Hyuga, Z. Phys. A 277 (1976) 291.

\bibitem{Hwang:1986pa} W.Y.P. Hwang, T.W. Donnelly, Phys. Rev. C 33 (1986) 1381.

\bibitem{Hwang:1985ie} W.Y.P. Hwang, G.E. Walker, Annals Phys. 159 (1985) 118.

\bibitem{Chuvilsky:1989} S.D. Kurgalin, Yu.M. Chuvilsky, Sov. J. Nucl. Phys. 49 (1989) 79.

\bibitem{Lykasov:2001} A.Yu. Illarionov, G.I.Lykasov, Phys. Rev. C 64 (2001) 044004.

\bibitem{Kukulin:2005} A. Faessler, V.I. Kukulin, M.A. Shikhalev, Annals Phys. 320 (2005) 71, nucl-th/0505026.

\bibitem{Uzikov:2009} D. Chiladze {\em et al.} Eur. Phys. J. A 40 (2009) 23.

\bibitem{krutov:2009} A.F. Krutov, V.E. Troitsky, Phys. Part. Nucl. 40 (2009) 136.

\bibitem{Azhgirey}
L.S. Azhgirey, N.P. Yudin,
\newblock Phys. Part. Nucl. 37, No. 4 (2006) 1011.

\bibitem{Bondarenko:2002zz}
S.G. Bondarenko, V.V. Burov, A.V. Molochkov, G.I. Smirnov, H.~Toki,
\newblock Prog. Part. Nucl. Phys. 48 (2002) 449, nucl-th/0203069.

\bibitem{Lacombe:1980dr}
M. Lacombe {\em et~al.},
\newblock Phys. Rev. C 21 (1980) 861.

\bibitem{Machleidt:1984xu}
R. Machleidt, K. Holinde,
\newblock Karlsruhe 1983 Proceedings, Few Body Problems In
Physics, Vol. 2 (1984) 79.

\bibitem{Crepinsek:1974pu}
L. Crepinsek, H. Oberhummer, W. Plessas, H. Zingl,
Acta Phys. Austriaca 39 (1974) 345.

\bibitem{Crepinsek:1975vn}
L. Crepinsek, C.B. Lang, H. Oberhummer, W. Plessas, H.F.K. Zingl,
\newblock Acta Phys. Austriaca 42 (1975) 139.

\bibitem{Haftel:1976zz}
M.I. Haftel,
\newblock Phys. Rev. C 14 (1976) 698.

\bibitem{Giraud:1979gw}
N. Giraud, Y. Avishai, C. Fayard, G.H. Lamot,
\newblock Phys. Rev. C 19 (1979) 465.

\bibitem{Mathelitsch:1981mr}
L.~Mathelitsch, W.~Plessas, M.~Schweiger,
\newblock Phys. Rev. C 26 (1982) 65.

\bibitem{Haidenbauer:1984dz}
J.~Haidenbauer, W.~Plessas,
\newblock Phys. Rev. C 30 (1984) 1822.

\bibitem{Haidenbauer:1986zza}
J. Haidenbauer, Y. Koike, W. Plessas,
\newblock Phys. Rev. C 33 (1986) 439.

\bibitem{Reid:1968sq}
R.V. Reid, Jr.,
\newblock Ann. Phys. 50 (1968) 411.

\bibitem{Ernst:1973zzb}
D.J. Ernst, C.M. Shakin, R.M. Thaler,
\newblock Phys. Rev. C 8 (1973) 46.

\bibitem{Rupp:1989sg}
G. Rupp, J.A. Tjon,
\newblock Phys. Rev. C 41 (1990) 472.

\bibitem{Schwarz:1980bc}
K.~Schwarz, J.~Frohlich, H.F.K. Zingl,
\newblock Acta Phys. Austriaca 53 (1981) 191.

\bibitem{Bondarenko:2004pn}
S.G. Bondarenko, V.V. Burov, K.Y. Kazakov, D.V. Shulga,
\newblock Phys. Part. Nucl. Lett. 1 (2004) 178, nucl-th/0402056v2.

\bibitem{Bondarenko:2008fp}
S.G. Bondarenko, V.V. Burov, W.-Y. Pauchy~Hwang, E.P. Rogochaya,
\newblock JETP Lett. 87 (2008) 653, 0804.3525v2 [nucl-th].

\bibitem{Bondarenko:2008ha}
S.G. Bondarenko, V.V. Burov, E.P. Rogochaya, Y.~Yanev,
\newblock 0806.4866 [nucl-th] (2008).

\bibitem{Bondarenko:2008mm}
S.G. Bondarenko, V.V. Burov, W.-Y. Pauchy~Hwang, E.P. Rogochaya,
\newblock Nucl. Phys. A 832 (2010) 233, nucl-th/0612071.

\bibitem{Noyes:1965ib}
H.P. Noyes,
\newblock Phys. Rev. Lett. 15 (1965) 538.

\bibitem{Kowalski:1965}
K.L. Kowalski,
\newblock Phys. Rev. Lett. 15 (1965) 798.

\bibitem{Salpeter:1951sz}
E.E. Salpeter, H.A. Bethe,
\newblock Phys. Rev. 84 (1951) 1232.

\bibitem{Kubis:1972zp}
J.J. Kubis,
\newblock Phys. Rev. D 6 (1972) 547.

\bibitem{Bethe:1949yr}
H.A. Bethe,
\newblock Phys. Rev. 76 (1949) 38.

\bibitem{Yamaguchi:1954mp}
Y. Yamaguchi,
\newblock Phys. Rev. 95 (1954) 1628;\\
Y. Yamaguchi, Y. Yamaguchi,
\newblock Phys. Rev. 95 (1954) 1635.

\bibitem{Dumbrajs:1983jd}
O.~Dumbrajs {\em et~al.},
\newblock Nucl. Phys. B 216 (1983) 277.

\bibitem{Lee:1969fy}
T.D. Lee, G.C. Wick,
\newblock Nucl. Phys. B 9 (1969) 209.

\bibitem{Rodning:1990zz}
N.L. Rodning, L.D. Knutson,
\newblock Phys. Rev. C 41 (1990) 898.

\bibitem{Honzawa:1991qn}
N. Honzawa, S. Ishida,
\newblock Phys. Rev. C 45 (1992) 47.

\bibitem{Fleischer:1975}
J. Fleischer, J.A. Tjon,
\newblock Nucl. Phys. B 84 (1975) 375.

\bibitem{Machleidt:2000ge}
R. Machleidt,
\newblock Phys. Rev. C 63 (2001) 024001.

\bibitem{Arndt:2007qn}
R.A. Arndt, W.J. Briscoe, I.I. Strakovsky, R.L. Workman,
\newblock Phys. Rev. C 76 (2007) 025209, 0706.2195v3 [nucl-th].

\bibitem{Bondarenko:1998fh}
S.G. Bondarenko, V.V. Burov, M. Beyer, S.M. Dorkin,
\newblock Phys. Rev. C 58 (1998) 3143.

\bibitem{Bondarenko:2006zq}
S.G. Bondarenko, V.V. Burov, E.P. Rogochaya, A.A. Goy,
\newblock Phys. Atom. Nucl. 70 (2007) 2054, nucl-th/0612071.

\bibitem{Bussiere:1981mv}
M. Bernheim {\em et al.},
\newblock Nucl. Phys. A 365 (1981) 349.

\bibitem{TurckChieze:1984fy}
S. Turck-Chieze {\em et al.},
\newblock Phys. Lett. B 142 (1984) 145.

\bibitem{Stoks:1993tb}
V.G.J. Stoks, R.A.M. Kompl, M.C.M. Rentmeester, J.J. de Swart,
\newblock Phys. Rev. C 48 (1993) 792.

\bibitem{Loiseau:1984eg}
B. Loiseau, L. Mathelitsch, W. Plessas, K. Schwarz,
\newblock Phys. Rev. C 32 (1985) 2165.



\end{thebibliography}
\end{document}